\documentclass[twocolumn,showpacs,preprintnumbers,amsmath,amssymb]{revtex4}

\usepackage{graphicx}
\usepackage{dcolumn}
\usepackage{bm}

\begin{document}


\title{Exploring the anomaly in the interaction cross section and matter radius of $^{23}$O } 

\author {R. Kanungo$^{1}$,  A. Prochazka$^{2,3}$, M. Uchida$^1$,  W. Horiuchi$^4$, G. Hagen$^{5,6}$,  T. Papenbrock$^{5,6}$,  C. Nociforo$^2$, T. Aumann$^{7,2}$,  D. Boutin$^3$,  D. Cortina-Gil$^8$,  B. Davids$^9$, M. Diakaki$^{10}$, F. Farinon$^{2,3}$, H. Geissel$^2$, R. Gernh\" auser$^{11}$, J. Gerl$^2$,   R. Janik$^{12}$, \O. Jensen$^{13}$, B. Jonson$^{14}$,  B. Kindler$^2$, R. Kn\" obel$^{2,3}$, R. Kr\" ucken$^{11}$, M. Lantz$^{14}$, H. Lenske$^3$, Y. Litvinov$^{2,15}$, B. Lommel$^2$, K. Mahata$^2$, P. Maierbeck$^{11}$, A. Musumarra$^{16}$,  T. Nilsson$^{14}$, C. Perro$^1$, C. Scheidenberger$^2$, B. Sitar$^{12}$, P. Strmen$^{12}$, B. Sun$^{2,17}$, Y. Suzuki$^{4,18}$, I. Szarka$^{11}$, I. Tanihata$^{19}$,  H. Weick$^2$, M. Winkler$^2$ }

\affiliation{$^1$Astronomy and Physics Department, Saint Mary's University, Halifax, NS B3H 3C3, Canada}
\affiliation{$^2$GSI Helmholtzzentrum f\"ur Schwerionenforschung, D-64291 Darmstadt, Germany}
\affiliation{$^ 3$Justus-Liebig University,  D-35392 Giessen, Germany}
\affiliation{$^{4}$RIKEN Nishina Center, Wako, Saitama 351-0918, Japan}
\affiliation{$^ 5$Physics Division, Oak Ridge National Laboratory, Oak Ridge, TN 37831, USA}
\affiliation{$^6$Department of Physics and Astronomy, University of Tennessee, Knoxville, TN 37996, USA}
\affiliation{$^7$Institut f\"ur Kernphysik, Technische Universit\"at Darmstadt, D-6489 Darmstadt, Germany}
\affiliation{$^8$Universidad de Santiago de Compostela, E-15706 Santiago de Compostella, Spain}
\affiliation{$^ 9$TRIUMF, Vancouver, BC V6T 2A3, Canada}
\affiliation{$^ {10}$National Technical University, Athens, Greece}
\affiliation{$^ {11}$Physik Department E12, Technische Universit\"at M\"unchen, D-85748 Garching, Germany}
\affiliation{$^ {12}$Faculty of Mathematics and Physics, Comenius University, 84215 Bratislava, Slovakia}
\affiliation{$^ {13}$Department of Physics and Center of Mathematics for Applications, University of Oslo, N-0316 Oslo, Norway}
\affiliation{$^ {14}$Chalmers University of Technology, G\"oteborg, SE 412-916, Sweden}
\affiliation{$^{15}$Max-Planck-Institut f\"ur Kernphysik, Saupfercheckweg 1, D-69117 Heidelberg, Germany}
\affiliation{$^{16}$Universita' di Catania, 95153 Catania, Italy }
\affiliation{$^{17}$School of Physics and Nuclear Energy Engineering, Beihang University, 100191 Beijing, PR China }
\affiliation{$^{18}$Department of Physics,  Niigata University, Niigata 950-2181, Japan}
\affiliation{$^{19}$RCNP, Osaka University, Mihogaoka, Ibaraki, Osaka 567 0047, Japan}

\date{\today}

\begin{abstract}
New measurements of the interaction cross sections of $^{22,23}$O at 900$A$ MeV performed at the GSI, Darmstadt  are reported that address the unsolved puzzle of the large cross section previously observed for $^{23}$O.
The matter radii for these oxygen isotopes extracted through a Glauber model analysis are in good agreement with the new predictions of 
the {\it ab initio} coupled-cluster theory reported here.
They are consistent with a $^{22}$O+neutron description of $^{23}$O as well. 

\end{abstract}

\pacs{25.60.-t, 25.60.Dz,  21.10.Gv, 21.60.De}
\maketitle

The nucleon magic numbers are the fundamental basis for the concept of nucleons being arranged in a shell structure. 
While the distributions of nucleons in stable nuclei are fairly well understood, the neutron-rich nuclei show signatures  of unconventional behaviour. Of particular interest is the region around the new magic number $N$=16 \cite{OZ01,KA09, HO09}, at the neutron drip-line that is a new benchmark point.
Here, the nucleus $^{23}$O plays a special role as a  large enhancement in the interaction cross section was observed for $^{22}$N, $^{23}$O and $^{24}$F \cite{OZ01}.  This leads to a large matter radius which was thought to reflect the formation of a neutron halo. However, the prediction by a core($^{22}$O)+(2$s_{1/2}$) neutron halo model was much lower than the data, and the matter distribution of $^{23}$O remained an unsolved puzzle being a challenge for all models of nuclear structure. 

The purpose of this Letter is to address this crucial issue, by measuring the interaction cross section and reliably extracting the root mean square radii,
in order to reach a conclusive understanding on $^{23}$O. 
The new observations resolve the existing anomaly, showing a smaller cross section consistent with both new {\it ab initio} model predictions presented here, and a  $^{22}$O core+ neutron description.

There are several indications of a sub-shell gap at $N$=14 in the oxygen isotopes. 
First, the 2$^+$ excitation energy of 3.199(8) MeV in $^{22}$O \cite{SO01, ST04} is large. Second, proton inelastic scattering of $^{22}$O \cite{BE06} revealed a small deformation parameter $\beta$=0.26$\pm$0.04.
This sub-shell gap suggests that the $^{23}$O ground state could have a large component of single-particle configuration with 2$s_{1/2}$ valence neutron. The nuclear and Coulomb breakup measurements of one-neutron removal from $^{23}$O reported the 2$s_{1/2}$ spectroscopic factors of 0.97$\pm$0.19 \cite{CO04, RO10} and 0.78$\pm$0.13  \cite{NO05} respectively. Significant yield of $^{22}$O excited state (3.2, 5.8 MeV) components were observed in the nuclear breakup. However, the $^{22}$O+n(2$s_{1/2}$) description in which the $^{22}$O core is considered identical to the bare $^{22}$O nucleus severely underpredicted the measured interaction cross section \cite{OZ01}. It was therefore proposed \cite{KA01} that the $^{22}$O core within $^{23}$O is possibly modified and enlarged compared to the bare $^{22}$O nucleus, giving rise to a larger interaction cross section.  The relatively narrow momentum distribution of the two-neutron removal fragment $^{21}$O from $^{23}$O  \cite{KA02} suggested that $^{23}$O might have some probability of two neutrons occupying the 2$s_{1/2}$ orbital.
However, this is not consistent with the reported 2$s_{1/2}$ spectroscopic factors.

The energy gap at $N$=14 in $^{23}$O was found to be 2.79(13) MeV from fragmentation of $^{26}$Ne,  populating the resonance in $^{23}$O at 45(2) keV above the $^{22}$O+n threshold \cite{SC07}. This was considered to be the 5/2$^+$ excited state in $^{23}$O. The higher lying resonance at 1.3 MeV above the neutron threshold observed in the $^{22}$O(d,p) reaction \cite{EL07} was understood as the 3/2$^+$ excited state that shows a 4.00(2) MeV gap between the 2$s_{1/2}$ and 1$d_{3/2}$ orbitals. 

Recently, the neutron knockout of $^{24}$O $\rightarrow ^{23}$O revealed a large two-neutron spectroscopic factor of 1.74$\pm$0.19 exhausting the s-orbital occupancy with no significant observable $d$- component establishing a spherical shell closure at $N$=16 \cite{KA09, JE11}. The neutron-unbound excited state of $^{24}$O at 4.72(11) MeV \cite{HO09}, considered to be the first 2$^+$ state, is also supportive of $^{24}$O being a doubly-magic nucleus. These findings are difficult to reconcile with the unusually large interaction cross section of $^{23}$O reported in \cite{OZ01}. Such an anomaly may point to unknown structure effects to which the breakup reactions may not be sensitive. The issue therefore is of utmost importance to be addressed since the neutron-rich oxygen isotopes form crucial benchmark points for understanding the evolution of shell structure in neutron-rich regions. It has been shown that three-body forces play an important role to define the drip-line of the oxygen isotopes \cite{OT10}.

We performed an experiment to measure the interaction cross sections of $^{22,23}$O at the fragment separator FRS at GSI \cite{FRS}.  The experiment layout is shown in Fig.1 of Ref.\cite{KA11}. The measurement is done by the method of transmission where the total interaction cross section for reactions in the target is given by  
$\sigma_I$=$(-1/t){\rm{ln}}(R_{in}/R_{out})$.
The transmission ratio is $R_{in}=N^f_{in}/N^i_{in}$ where $N^i_{in}$ and $N^f_{in}$ are the number of $^A$O before and after the target, respectively. $R_{out}$ is the same but for an empty target and $t$ is the number of target nuclei per unit area.

The $^{A}$O nuclei were produced from the fragmentation of a 1$A$ GeV $^{48}$Ca beam interacting with a 6.3 g/cm$^2$ thick Be target. The fragmentation products were separated and identified using the first half of the FRS, where plastic scintillator detectors placed at the two dispersive foci, F1 and F2, measured the time-of-flight (TOF).  Precise beam position measured using time projection chambers (TPC) detectors placed at F2 and the magnetic rigidity of the FRS provide information on the mass to charge ratio ($A/Q$) of the fragments. A multi-sampling ionization chamber (MUSIC) placed at F2 measured the energy-loss (with a 1$\sigma$ resolution of $\sim$ 3\%) providing the $Z$ identification of the $^{A}$O beam. 

A  4.046 g/cm$^2$ thick carbon reaction target was placed at F2. The second half of the fragment separator consists of a dispersive focus at F3 and an achromatic focus at F4. The ion optics mode was selected to match the dispersion of the first half with the second half. The magnetic rigidity of the second half was set to transport the unreacted $^A$O to the final focus (F4). Here products were identified in $A/Q$ using magnetic rigidity, TOF between plastic scintillators at F2 and F4, and position measurement using TPC detectors placed at F4. Two MUSIC  detectors were placed at F4 to measure the energy-loss of the products after the reaction target. Events that were observed to be consistent with $Z$=8 in either MUSIC were counted as unreacted events.  To account for losses occurring due to reactions in other materials in the setup, data were also collected with an empty target frame. The position and angle of the incident beam before the target at F2 were restricted from beam tracking measurements such that transmission from F2 to F4 was a constant within the selected phase space. 

The measured interaction cross section at $\sim$900$A$ MeV is shown in Fig. 1 as a function of the mass number of the oxygen isotopes with the filled circles showing the present data from Table 1.  The open squares are the data of Ref.\cite{OZ01}. It is seen that the present data of $^{23}$O is smaller than that reported earlier. The total uncertainties shown are dominated by statistics. The uncertainty from  contamination of the incoming beam (for $Z$ and $A$) was at most $\sim$ 0.9\%, and that from the target thickness was 1.2 \%. Transmission uncertainty was $\sim$ 1.5\%.  The cross section for $^{22}$O is also slightly smaller than previously reported. The cross section for the isotope $^{28}$Ne extracted from the same data results in 1271$\pm$35 mb which agrees within uncertainty with the value of 1244$\pm$44 mb reported in Ref.\cite{OZA01}.
The interaction cross section of $^{23}$O reported here is $\sim$ 8-9\% larger than $^{22}$O, which may not be sufficient to be classified as a one-neutron halo ($^{11}$Be, a halo,  has $\sim$ 16\% larger cross-section than $^{10}$Be). 
This is consistent with the fact that the one-neutron separation energy of $^{23}$O, $S_n$=2.7 MeV is quite large which inhibits the tunnelling of the wavefunction into the classically forbidden region to form a halo. 

\begin{figure}
\includegraphics[width=5cm, height=5cm]{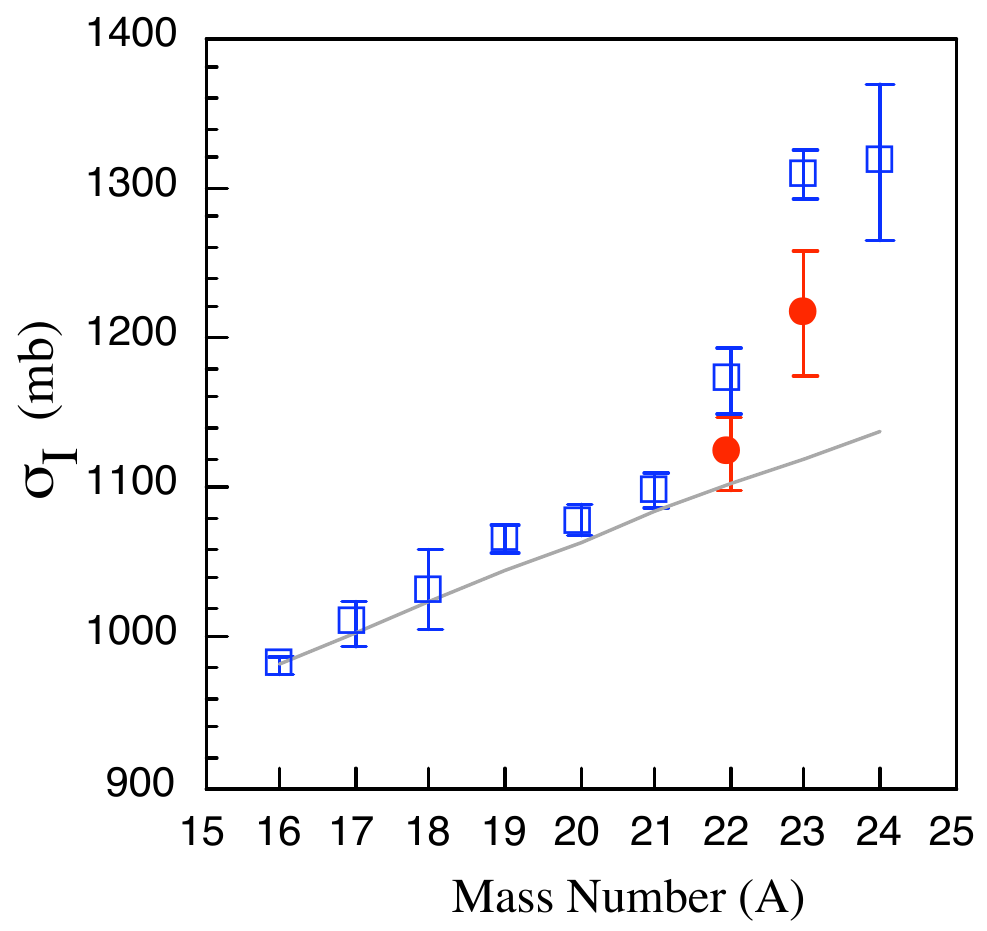} 
\caption{\label{fig} The interaction cross section of $^A$O+C as a function of the mass number. The circles represent the present data, and the squares are from Ref.\cite{OZ01}. The line shows the $A^{1/3}$ dependence normalized to $^{16}$O.}
\end{figure}

The root mean square ($r.m.s.$) matter radius is extracted by interpreting the data in the framework of the Glauber model including the higher order terms that are missing in the usual optical  approximation \cite{AB00, HO07}. The matter density was considered to be a Fermi density of the form $\rho(r) = \rho_0/(1+{\rm{exp}}((r-R)/a))$ where $R = r_0A^{1/3}$. The calculated cross sections for different values of radius ($r_0\sim 0.8-1.4$ fm ) and diffuseness ($a\sim 0.3-0.7$ fm) parameters are shown by the different filled points in Figs.2a and 2b. Calculations using separate neutron and proton Fermi functions were found to yield similar $r.m.s.$ radii. The $r.m.s.$ matter radius that can reproduce the experimental cross section is found to be 2.75$\pm$0.15 fm for $^{22}$O while for $^{23}$O a radius of 2.95$\pm$ 0.23 fm is extracted (Table 1, shown by the horizontal arrows in Figs. 2a and b). To investigate any model dependence we also extract the $r.m.s.$ radii with a harmonic oscillator form of density \cite{OZA01}. The radii extracted shown by solid line (open points) in Figs. 2a and 2b are consistent with those using Fermi density (Table 1). Since the oscillator width is the only one parameter here, the uncertainty of the radius is found to be slightly smaller. 

A sub-shell closure at $N$=14 has been discussed for $^{22}$O which allows us in a simplistic model to describe $^{23}$O being composed of a $^{22}$O core+neutron and we interpret next the interaction cross section in a few-body Glauber model framework \cite{AB03}.
The core $^{22}$O is considered to have the same Fermi density profile as the bare $^{22}$O nucleus mentioned above. The wavefunctions of the valence neutron are calculated with a Woods-Saxon bound state potential ($r_0$=1.27 fm and $a$=0.67 fm) where the depth is varied to reproduce the effective neutron separation energy. The dashed line in Fig.2c shows the calculated cross sections with the valence neutron in the 2$s_{1/2}$ orbital. The cross section shown is calculated as a function of the rms matter radius of $^{22}$O (from Table 1). The horizontal shaded area is the measured interaction cross section of $^{23}$O. The overlap of the calculated values with this shaded region shows consistency with data. It is seen therefore, that $^{23}$O can be described by a $^{22}$O core+2$s_{1/2}$ neutron. The dotted line shows a similar calculation but for the neutron in the 1d$_{5/2}$ orbital and with the core $^{22}$O in its 2$^+$ excited state. The density of $^{22}$O in its 2$^+$ state was assumed to be the same as the ground state.  

\begin{figure}
\includegraphics[width=5cm, height=12cm]{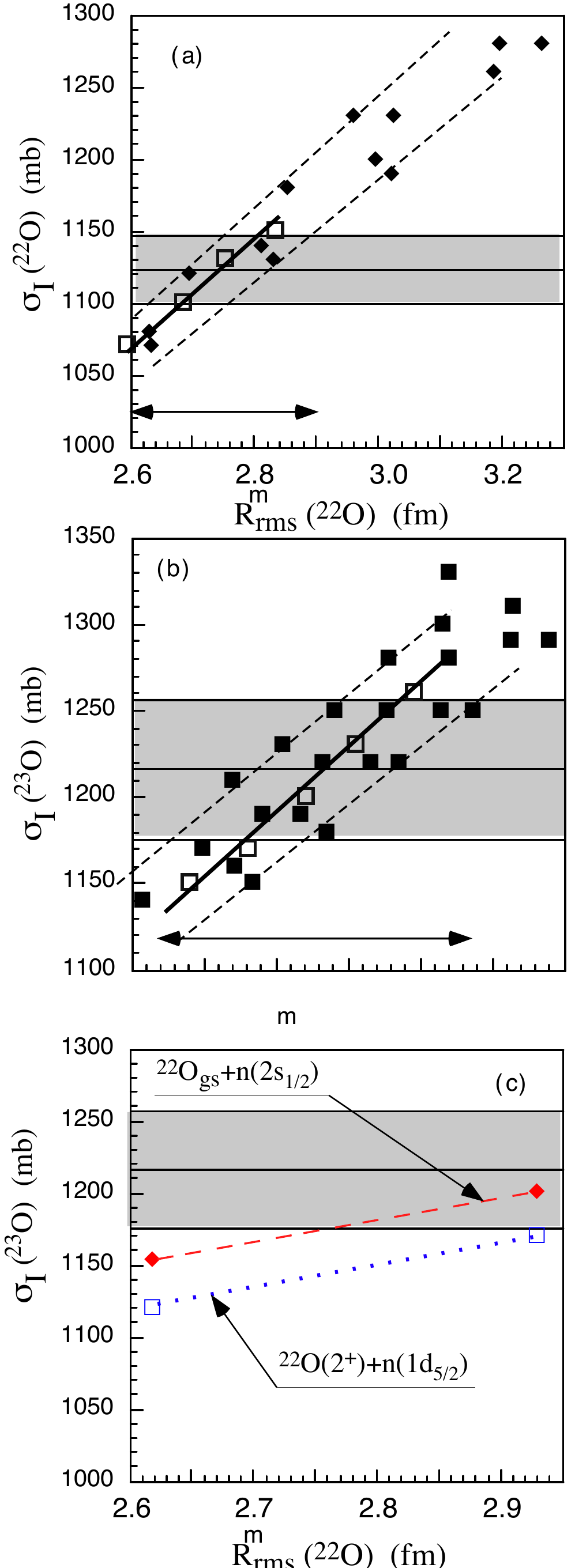} 
\caption{\label{fig} The interaction cross section data for (a) $^{22}$O+C and (b) $^{23}$O+C (shaded regions). The squares are Glauber model calculations with different Fermi density parameters $r_0$ and $a$ shown as a function of R$^m_{rms}(^A$O). The dotted lines show a guide to the eye for determining the limits of R$^m_{rms}$ consistent with the data. The solid line are calculations using harmonic oscillator density. (c) The interaction cross section data for $^{23}$O+C (shaded region). The dashed (dotted) line is $^{22}$O+n few-body Glauber model calculation for 2s$_{1/2}$ (1d$_{5/2}$) orbitals as a function of different $^{22}$O
$r.m.s.$ radii within uncertainty of the value quoted in Table 1. }
\end{figure}

To gain a better understanding, we perform
{\it ab initio} coupled-cluster (CC) ~\cite{CC} computations. This method is
uniquely suited for describing nuclei with closed neutron and
proton subshells and their neighbors. Within particle-removed CC theory, 
the ground state of  $^{23}$O
is described as a superposition of 1$h$ and
1$p$Ð2$h$ excitations on top of the correlated ground state of 
$^{24}$O \cite{long}. We employ a
low-momentum version of the nucleon-nucleon interaction from chiral
effective field theory ~\cite{EM} that results from a similarity
renormalization group transformation~\cite{SRG} and is characterized
by a momentum cutoff $\lambda$.  We work with the intrinsic
Hamiltonian, i.e. the kinetic energy of the center of mass is
subtracted from the total kinetic energy. As a result, the
coupled-cluster wave function factorizes with a high degree of
accuracy into a product of an intrinsic wave function and a Gaussian
for the center of mass~\cite{HPD}. In this framework we compute the density 
of the $^{23}$O ground state. The intrinsic density, i.e. the density with
respect to the center of mass, results from a deconvolution with
respect to the Gaussian center-of-mass wave function. It
enters the computation of the interaction cross section and the matter radii.
Our model space consisted of 30 bound and continuum Woods-Saxon orbitals 
for the neutron, $l$=0 and $l$=2 partial waves, and 17 major oscillator shells for the
remaining neutron partial waves and the protons.

The computed point matter radii are in good agreement with our data (Fig.3a). 
Smaller values of the momentum cutoff lead to smaller radii.
The relative uncertainties in the experimental radii
are larger than those of the interaction cross sections due to the
uncertainties in the Fermi density profiles. 
We also compare the cross sections calculated with density distributions from the coupled cluster theory
in Fig. 3b. 
The cross sections with $\lambda=4.0$ \& $3.8$~fm$^{-1}$ are in good agreement with the data,
while that with $\lambda$=$3.6$~fm$^{-1}$ is slightly below the $1\sigma$ error. The variation of the radii with the cutoff provides an estimate of the contributions of neglected  short-ranged three-nucleon forces.
The relative isotopic differences in radii from $^{21-24}$O, depend only very weakly on the cutoff, suggesting it to probably have a weak dependence on three-nucleon forces.
The results with densities from Ref.\cite{AB09} which were generated by a Slater determinant in a mean field potential are shown by the cross marks  in (Fig. 3). The open circles (Fig. 3) show results with densities from a Skyrme-Hartree-Fock potential \cite{BA01}.

\begin{figure}
\includegraphics[width=8.5cm, height=4.5cm]{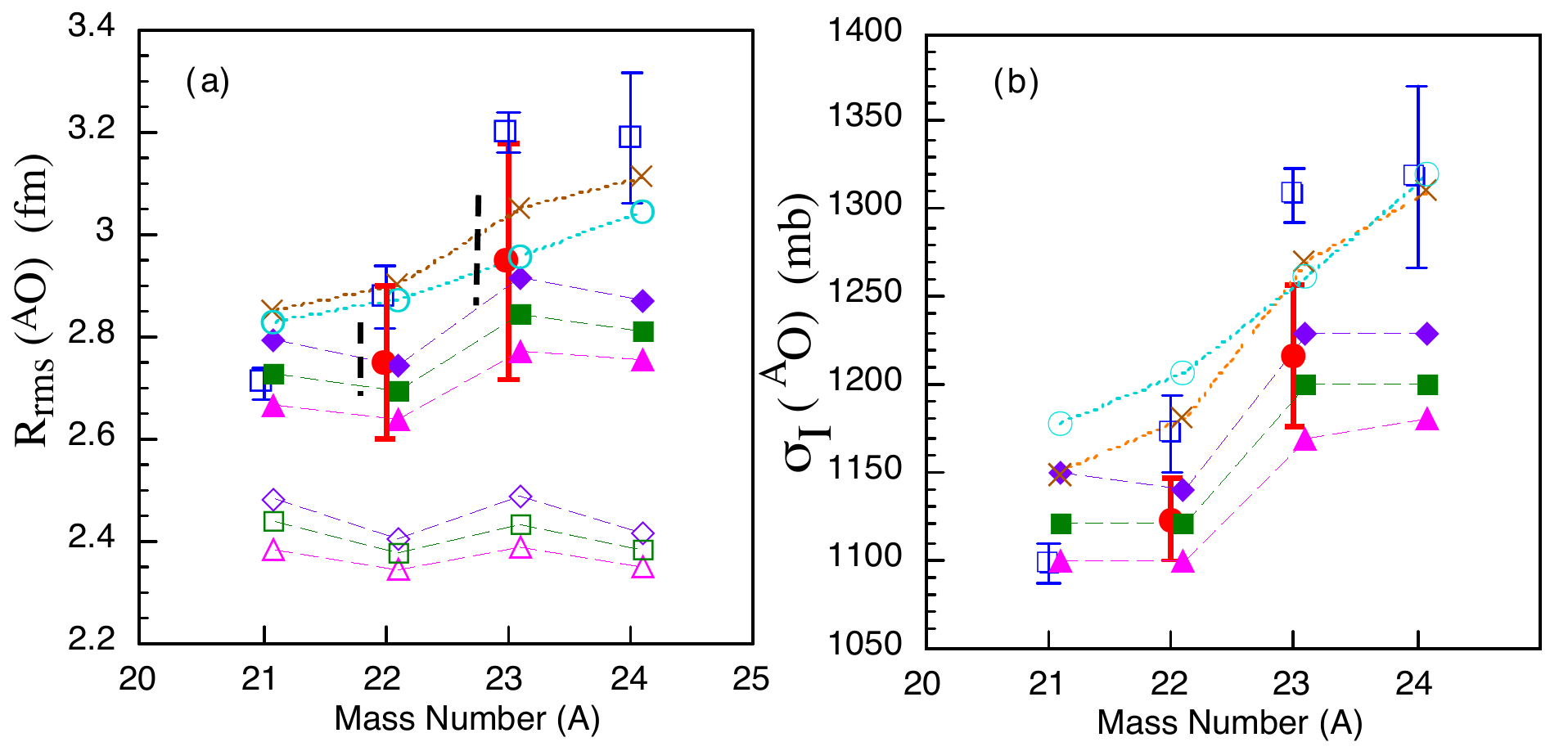} 
\caption{\label{fig}  (a) The red filled circles show R$^{m(Fermi)}_{rms}$, dashed vertical line is  R$^{m(HO)}_{rms}$ and the blue open squares are R$^m_{rms}$ from \cite{OZ01}. The diamonds /squares / triangles are coupled-cluster calculations with cut-off parameter = 4.0 / 3.8 / 3.6 fm$^{-1}$ where the filled points are R$^m_{rms}$ and the open ones are proton $r.m.s.$ radii respectively. (b) The interaction cross section of $^A$O+C as a function of the mass number. The red filled circles are present data, the squares are data from Ref.\cite{OZ01}.  The diamonds /squares / triangles are CC calculations with cut-off parameter = 4.0 / 3.8 / 3.6 fm$^{-1}$. The cross marks (open circles) show results from Ref.\cite{AB09} (Ref.\cite{BA01}). }
\end{figure}

In conclusion, this work reports new measurements of the interaction cross sections of $^{22,23}$O at 900$A$ MeV.
The new data for $^{23}$O is consistent, within experimental errors, with a model of $^{22}$O core + valence neutron in the 2$s_{1/2}$ orbital 
thereby addressing the existing anomaly in its structure.  Beyond this simplistic model, new coupled-cluster calculations reported here, show excellent agreement with the present data. This shows the significant advancement jointly by experiment and {\it ab initio} theories in reaching a conclusive understanding on the matter distribution in $^{23}$O. The radii extracted are shown to be consistent for the different parametric density forms. 
The results show  growth of neutron skin from $^{21}$O to $^{23}$O, but a large enhancement characteristic of a halo is not observed for $^{23}$O.

\begin{table}
\caption{\label{tab:table1} Measured interaction cross sections and the root mean square point matter radii  ($R^m_{rms}$) for
$^{22-23}$O using Fermi (Fermi) and Harmonic Oscillator (HO) densities.}
\begin{ruledtabular}
\begin{tabular}{llllll}
Isotope & $\sigma_I(\Delta \sigma)$&$\Delta \sigma$(Stat.)&$\Delta \sigma$(Syst.)&$R^{m(Fermi)}_{rms}$ & $R^{m(HO)}_{rms}$ \\
&(mb)&(mb)&(mb)&(fm)&(fm) \\
\hline
$^{22}$O &1123(24)&18.5&15.3& 2.75(0.15) & 2.75(0.07)\\
$^{23}$O &1216(41)&33.1&24.7& 2.95(0.23)& 2.97(0.11) \\

\end{tabular}
\end{ruledtabular}
\end{table}


The authors are thankful for the support of the GSI accelerator staff and the FRS technical staff for an efficient running of the experiment. The support from NSERC for this work is gratefully acknowledged.  This work was supported by the BMBF under contract 06MT238, by the DFG cluster of excellence {\it Origin and Structure of the Universe}. This work was supported by the 
U.S. Department of Energy, Grants No. DE-FG02-96ER40963 (University of Tennessee), and
No. DE-FC02-07ER41457 (UNEDF SciDAC). This research used computational resources of the National Center for Computational Sciences. W. H. is supported by the Special Postdoctoral Researchers Program of RIKEN. We thank B.A. Brown for the providing us the values of Ref.\cite{BA01}. 


\end{document}